\documentclass[3p,times,procedia]{elsarticle}

\usepackage{ecrc}
\usepackage{amsmath}
\usepackage{subfigure}

\volume{00}

\firstpage{1}

\journalname{Physics Procedia}

\runauth{S. Schopferer et al.}

\jid{phpro}

\jnltitlelogo{Physics Procedia}

\CopyrightLine{2011}{Elsevier BV. Selection and/or peer-review under responsibility of the organizing committee for TIPP 2011.\\
NOTICE: This is the author's version of a work accepted for publication by Elsevier. Changes resulting from the publishing process, including peer review, editing, corrections, structural formatting and other quality control mechanisms, may not be reflected in this document. Changes may have been made to this work since it was submitted for publication.}

\usepackage{amssymb}

\usepackage[figuresright]{rotating}

\begin{document}

\begin{frontmatter}



\dochead{TIPP 2011 -  Technology and Instrumentation for Particle Physics 2011}

\title{The GANDALF 128-Channel Time-to-Digital Converter}


\author[]{M.~B\"uchele}
\author[]{H.~Fischer}
\author[]{F.~Herrmann}
\author[]{K.~K\"onigsmann}
\author[]{C.~Schill}

\author[]{S.~Schopferer\corref{cor1}}
\ead{sebastian.schopferer@cern.ch}

\cortext[cor1]{Corresponding author}
\address{Universit\"at Freiburg, Physikalisches Institut, Hermann-Herder-Str. 3, 79104 Freiburg, Germany}

\begin{abstract}
The GANDALF 6U-VME64x/VXS module has been designed to cope with a variety of readout tasks in high energy and nuclear physics experiments, in particular the COMPASS experiment at CERN.
The exchangeable mezzanine cards allow for an employment of the system in very different applications such as analog-to-digital or time-to-digital conversions, coincidence matrix formation, fast pattern recognition or fast trigger generation. 
Based on this platform, we present a 128-channel TDC which is implemented in a single Xilinx Virtex-5 FPGA using a shifted clock sampling method. In this concept each input signal is continuously sampled by 16 flip-flops using equidistant phase-shifted clocks. Compared to previous FPGA designs, usually based on delay lines and comprising few TDC channels with resolutions in the order of 10\:ps, our design permits the implementation of a large number of TDC channels with a resolution of 64\:ps in a single FPGA. Predictable placement of logic components and uniform routing inside the FPGA fabric is a particular challenge of this design. We present measurement results for the time resolution and the nonlinearity of the TDC readout system.
\end{abstract}


\end{frontmatter}




\section{The GANDALF Framework}

\begin{figure}[!t]
\centering
\includegraphics[width=1\textwidth]{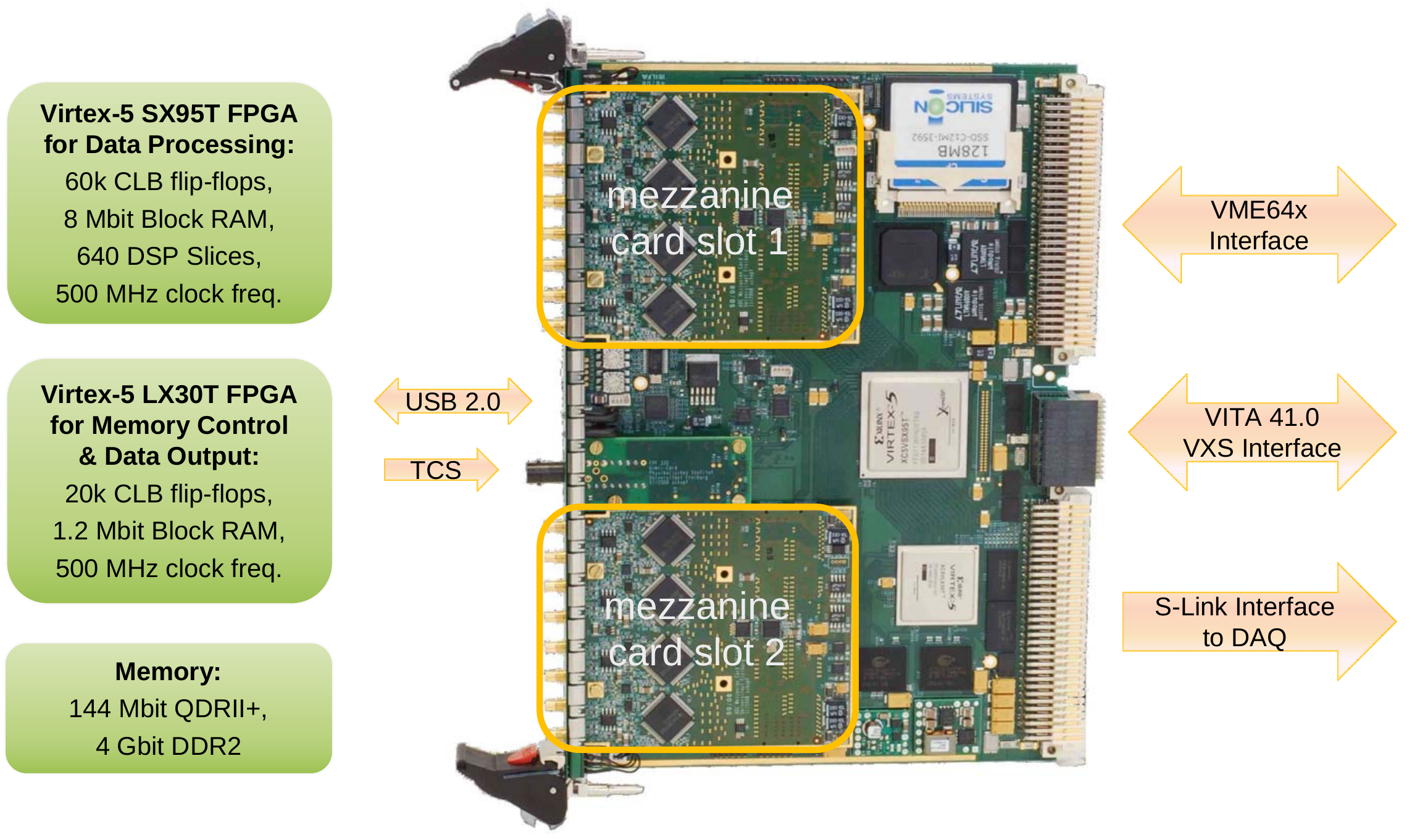}
\caption{Picture of the GANDALF carrier board equipped with two ADC mezzanine cards. The center mezzanine card hosts an optical receiver for the COMPASS trigger and clock distribution system (TCS). The VME64x interface is used for configuration and monitoring of the board, data is sent to the data acquisition system (DAQ) via the S-Link or the USB interface, and the VXS interface is used for inter-board communication.}
\label{fig_gandalf1}
\end{figure}

GANDALF \cite{nim2010,ieee2011} is a 6U-VME64x/VXS carrier board which can host two custom mezzanine cards (Fig. \ref{fig_gandalf1}). It has been designed to cope with a variety of readout tasks in high energy and nuclear physics experiments. 
Depending on the requirements of the desired application, the system can be equipped with different types of mezzanine cards.
Currently three types of mezzanine cards are available: 8-channel ADC cards, 64-channel LVDS input cards and 64-channel LVDS output cards. Presently under development is a card with high-speed optical interfaces for data transfer to/from remote detector frontend modules.

\begin{figure}[!b]
	\centering
		\includegraphics[width=0.495\textwidth]{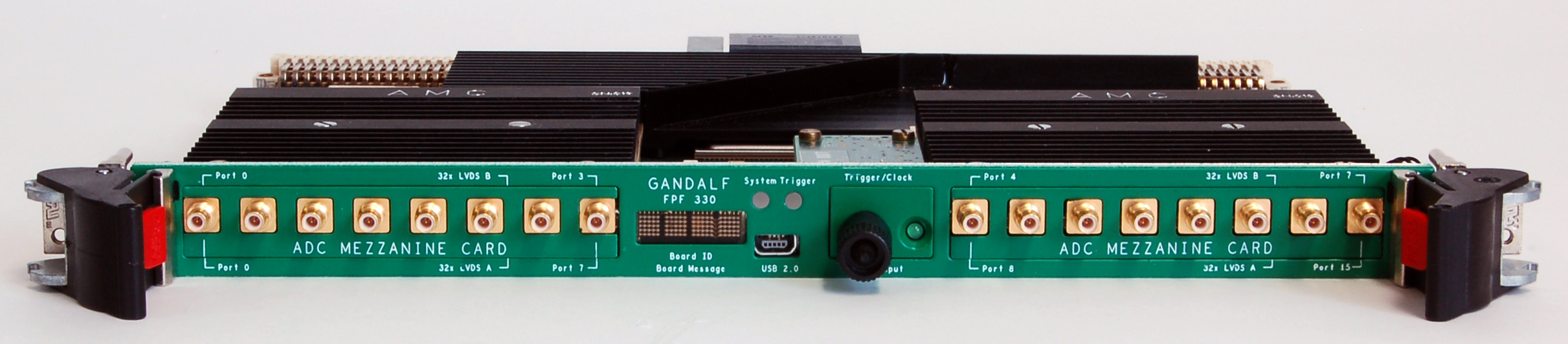}
		\hfill
		\includegraphics[width=0.495\textwidth]{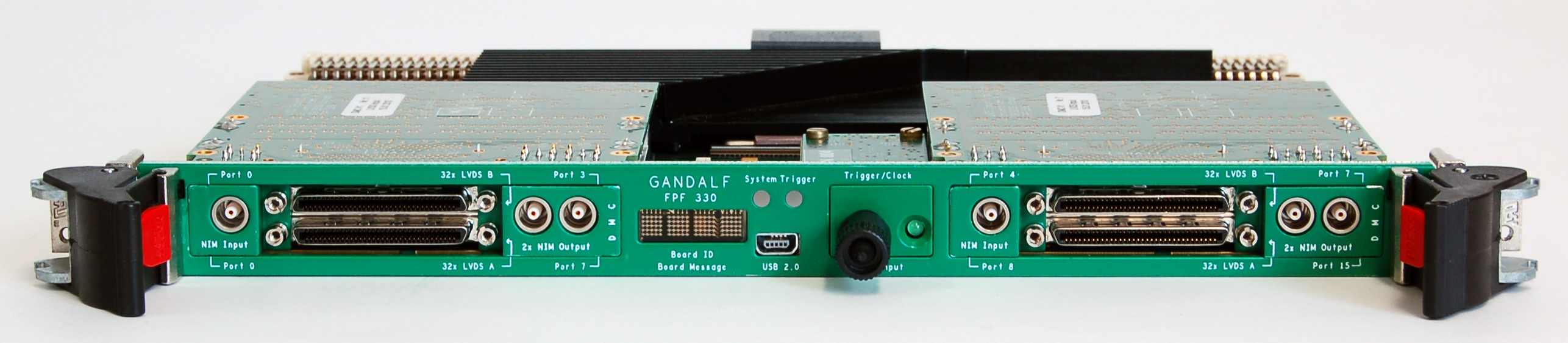}
	\caption{Two possible configurations of the GANDALF module. Left (with ADC cards): transient-analyzer with real-time pulse shape analysis and online feature extraction. Right (with LVDS input cards): 128-channel time-to-digital converter.}
	\label{fig_gandalf2}
\end{figure}

The mainboard comprises two Xilinx Virtex-5 FPGAs. The main FPGA (Virtex-5 SX95T) includes a large number of DSP slices which are used for fast signal processing when GANDALF is operated with ADC cards (Fig. \ref{fig_gandalf2}, left) in transient-analyzer mode \cite{florian,sebastian}. For digital I/O applications the board is equipped with the LVDS I/O cards (Fig. \ref{fig_gandalf2}, right). The differential signals are routed directly to the main FPGA, where the desired logic is implemented. Several applications like a 128-channel scaler, a 64-channel mean-timer with subsequent coincidence matrix \cite{meantimer} and a pattern generator have been successfully implemented so far.

Fast and deep memory extensions of 144-Mbit QDRII+ and 4-Gbit DDR2 RAM are connected to a second FPGA (Virtex-5 LX30T). Both FPGAs are linked to each other by eight bidirectional high-speed Aurora lanes with a total bandwidth of 25 Gbit/s per direction. A dead-time free data output can be realized by dedicated backplane link cards, following the 160 MByte/s S-Link \cite{SLINK} or the Ethernet protocol. Alternatively a data readout is possible by using the VME64x bus in block read mode or the USB2.0 port on the front panel. The VME and USB protocols are handled by a Xilinx CoolRunner-II CPLD.

\subsection{The LVDS Input and Output Cards}
The LVDS input card provides 64 differential inputs via two VHDCI\footnote{Very High Density Cable Interconnect} connectors. The signals are transferred by differential buffers\footnote{On Semiconductor NB4N855S} converting signal levels and protecting the FPGA from short circuits and electrostatic discharges. The jitter of the signal path including the buffers and the FPGA inputs is below 20\:ps RMS. Additionally, one NIM input and two NIM outputs are available via LEMO connectors, e.g. for gating or triggering purposes. With the same PCB a LVDS output card can also be assembled by different placement of the components.

\subsection{The Virtex-5 Architecture}
The Xilinx Virtex-5 is a powerful FPGA built on a 65-nm copper CMOS process technology \cite{xilinx}. The SX95T contains 160 x 46 Configurable Logic Blocks (CLBs) which are made up of two slices each. Every slice contains four function generators (6-input look-up tables), four storage elements (D-type flip-flops or latches), fast carry logic, large multiplexers and connections to a switch matrix to access general routing resources.
Furthermore the FPGA contains 244 blocks of 36-Kbit RAM (configurable as dual-port RAM or FIFOs) and 640 DSP48E slices with a 25 x 18 two's complement multiplier and a 48-bit arithmetic logic unit usable as adder, subtracter, accumulator or bit-wise logic unit.
Six clock management tiles with digital clock managers and PLLs are available for input jitter filtering, frequency synthesis, clock division and clock phase shifting.
The LX30T is a smaller FPGA model based on the same architecture, providing approx. 30\% of the logic resources and 15\% of the memory resources compared to the SX95T.

\section{The GANDALF Time-to-Digital Converter}

This section describes the implementation of 128 TDC channels inside the Virtex-5 FPGA located on the GANDALF board. The design objectives for this project were based on the requirements of high-rate particle physics experiments. The time resolution is required to be better than 100 ps RMS for precise tracking and time-of-flight measurements. The TDC has to be multi-hit capable with a deep hit-buffer and a programmable trigger window to select hits in the region of interest around a trigger. A dead-time free digitization has to be guaranteed even for bursts of many consecutive hits and triggers.

\subsection{TDC Concepts}
\begin{figure}
	\centering
		\subfigure[delayed data sampling]{\label{fig_tdcconcept-a}\includegraphics[width=0.495\textwidth]{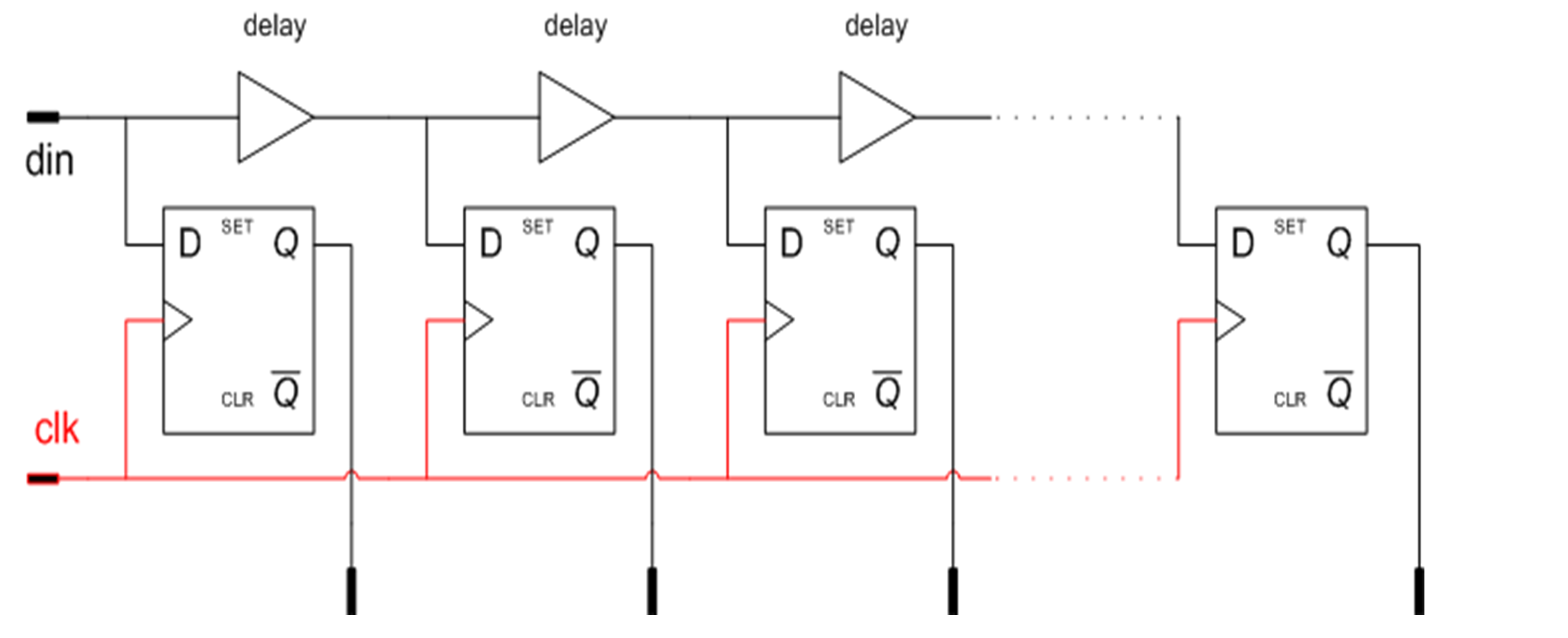}}
		\hfill
		\subfigure[shifted clock sampling]{\label{fig_tdcconcept-b}\includegraphics[width=0.49\textwidth]{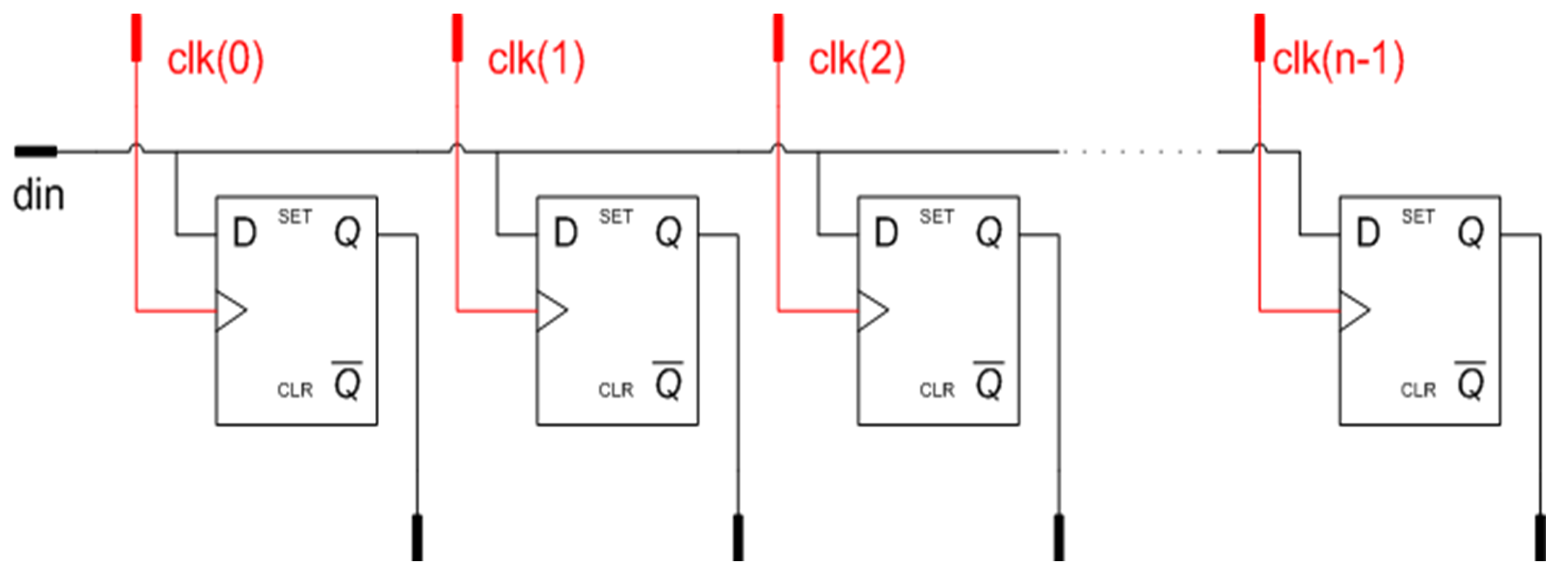}}
	\caption{Two possible TDC concepts for implementation in a FPGA.}
	\label{fig_tdcconcept}
\end{figure}

There are different concepts to implement a TDC in a FPGA. A trivial TDC would just sample the data signal with one flip-flop, resulting in a TDC bin width of $1/f_{clk}$. Since the clock frequency in a FPGA is limited to $f_{clk} \approx$ 500 MHz, one has to subdivide the clock period to achieve the desired resolution. 
Fig.\:\ref{fig_tdcconcept} shows two possible concepts: for the delayed data sampling (DDS, Fig.\:\ref{fig_tdcconcept-a}) the input signal is routed through a tapped delay line and the delayed signals from the taps are sampled by flip-flops with one common clock. This results in a bit pattern depending on the propagation time of the signal through the delay line until the next rising edge of the sampling clock. For the shifted clock sampling (SCS, Fig.\:\ref{fig_tdcconcept-b}) the input signal is routed with minimum skew to a number of flip-flops which are clocked by a set of $n$ equidistant phase-shifted clocks $\mathrm{clk}(i)$ with $i=0,1,\ldots,n-1$.

Both concepts have of course their pros and cons. While the DDS uses only one common clock, which makes it easy to further process the sampled data, the SCS starts from a set of different clock domains which have to be synchronized first. The main drawback of the DDS is the allocation of the delay elements in an FPGA. Various routing resources with different propagation delays (like the carry lines or the general routing matrix) are available but the delays are non-uniform, so every TDC channel has to be calibrated. With the dedicated carry-chains, high-resolution TDCs have been implemented in FPGAs so far, but the logic consumption for 128 TDC channels would exceed by far the device resources.

\subsection{Shifted Clock Sampling}
\label{label_scs}
\begin{figure}[!t]
	\centering
		\includegraphics[width=0.95\textwidth]{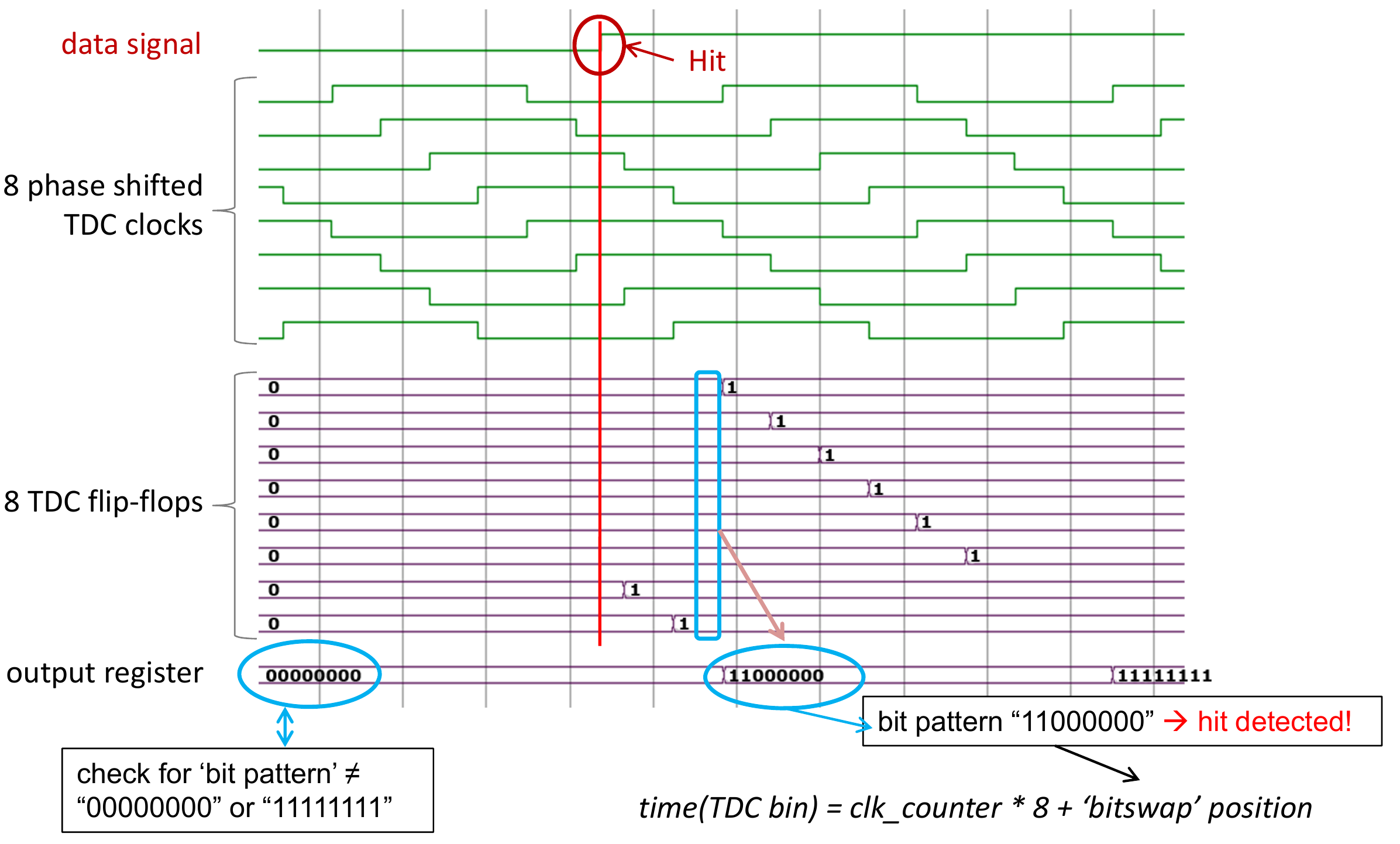}
	\caption{Signal timing diagram of the 8-bin shifted clock sampling TDC. For a detailed explanation see section \ref{label_scs}.}
	\label{fig_simu}
\end{figure}

The implementation presented in this article is based on the SCS method. Fig. \ref{fig_simu} shows a simplified signal timing diagram of an 8-bin TDC. The first line represents the data input signal with a hit (rising edge). The 8 TDC clocks $\mathrm{clk}(i)$ are equally phase-shifted by $\Delta\phi(i) = i \cdot 2 \pi / 8$. They clock the 8 TDC flip-flops which are all connected to the same data input signal. The flip-flop with a rising clock edge right after the hit (in this example the next-to-last one) is the first to sample the new value ('1'). The other flip-flops are following shortly after. Once every clock period, the values from all flip-flops are copied to an output register. The hit searching algorithm tests the output register for bit patterns $\neq$ "00000000" or "11111111". If a pattern with a 'bitswap' (change from 0s to 1s, or from 1s to 0s) is found, the bitswap position together with the value of the clock counter contains the time information of the hit. The hit searching algorithm can be configured to be leading and/or trailing edge sensitive.

\begin{figure}[!t] 
	\centering
		\includegraphics[width=0.95\textwidth]{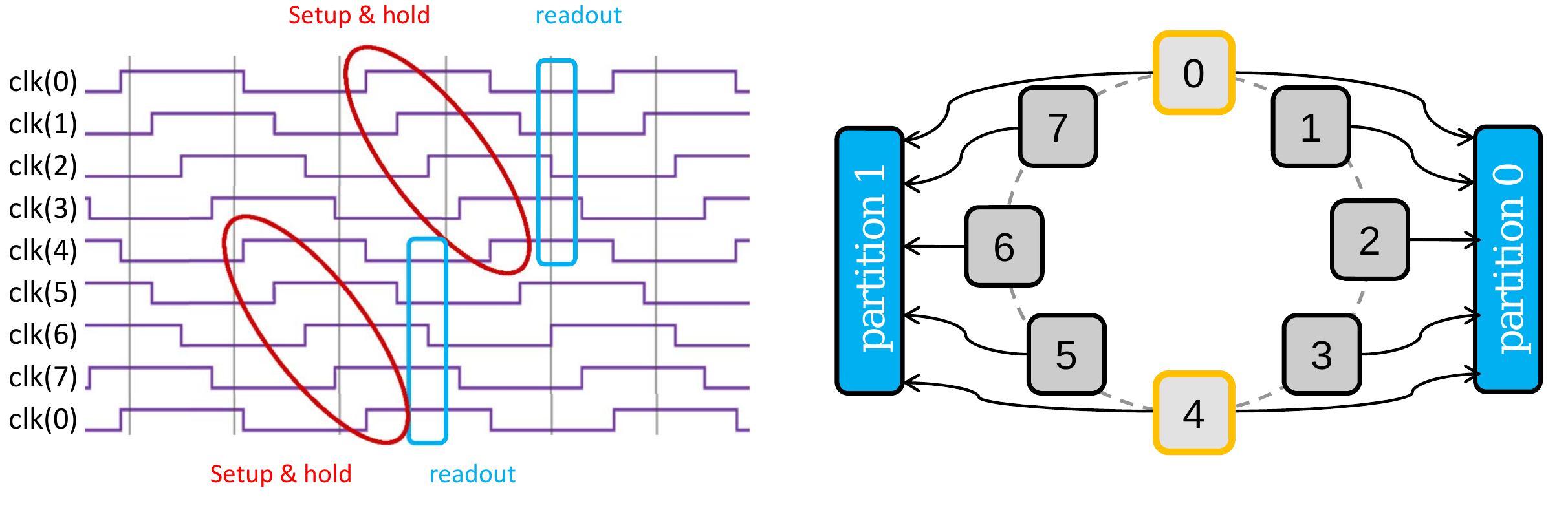}
	\caption{Readout of the 8-bin TDC. To meet the setup \& hold requirements, two partitions are introduced. Half of the TDC flip-flops are read out at a time.}
	\label{fig_part}
\end{figure}

Due to setup \& hold requirements the TDC flip-flops cannot be read out simultaneously like it is shown in Fig. \ref{fig_simu}. To synchronize the flip-flop outputs, the different clock domains are merged in a two-stage process (Fig. \ref{fig_part}). Two 'partitions' are introduced, each reading half of the flip-flops. Because a bitswap can only be detected within a partition, the flip-flops at the partition borders (no. 0 and 4 in the figure) are read from both partitions to avoid the loss of hits that might occur on these borders.

\begin{figure}[!t]
	\centering
		\includegraphics[width=0.48\textwidth]{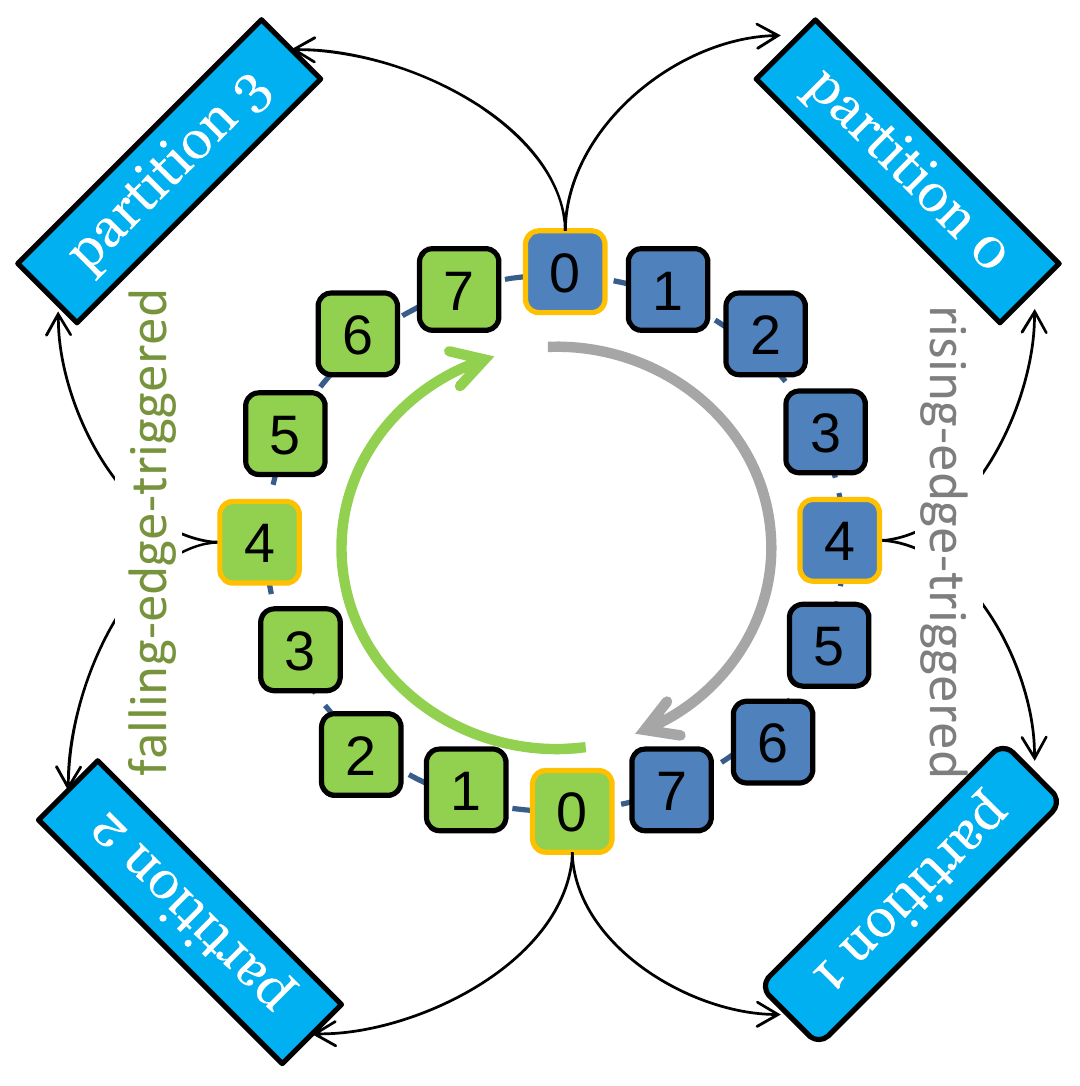}
	\caption{The 16-bin TDC design is based on 4 partitions. The squares illustrate the TDC flip-flops and the numbers denote the index $i$ of the corresponding clock. Flip-flops that are triggered by a rising clock edge are drawn in blue, falling edge triggered flip-flops are drawn in green.}
	\label{fig_16part}
\end{figure}

For the sake of clarity the example above describes an 8-bin TDC, however, the final design (Fig. \ref{fig_16part}) uses 16 TDC flip-flops, hence dividing the bin width by another factor of two. The 8 TDC clocks $\mathrm{clk}(i)$ ($i=0\ldots7$) are phase-shifted by $\Delta\phi(i) = i \cdot 2 \pi / 16$, therefore spanning half a clock period. The first 8 flip-flops are rising-edge triggered, the others are falling-edge triggered by locally inverting the clocks in the corresponding slices. The synchronization of the clock domains is performed by using 4 partitions.

\subsection{128-Channel TDC Design}
\begin{figure}[!t]
	\centering
		\includegraphics[width=0.95\textwidth]{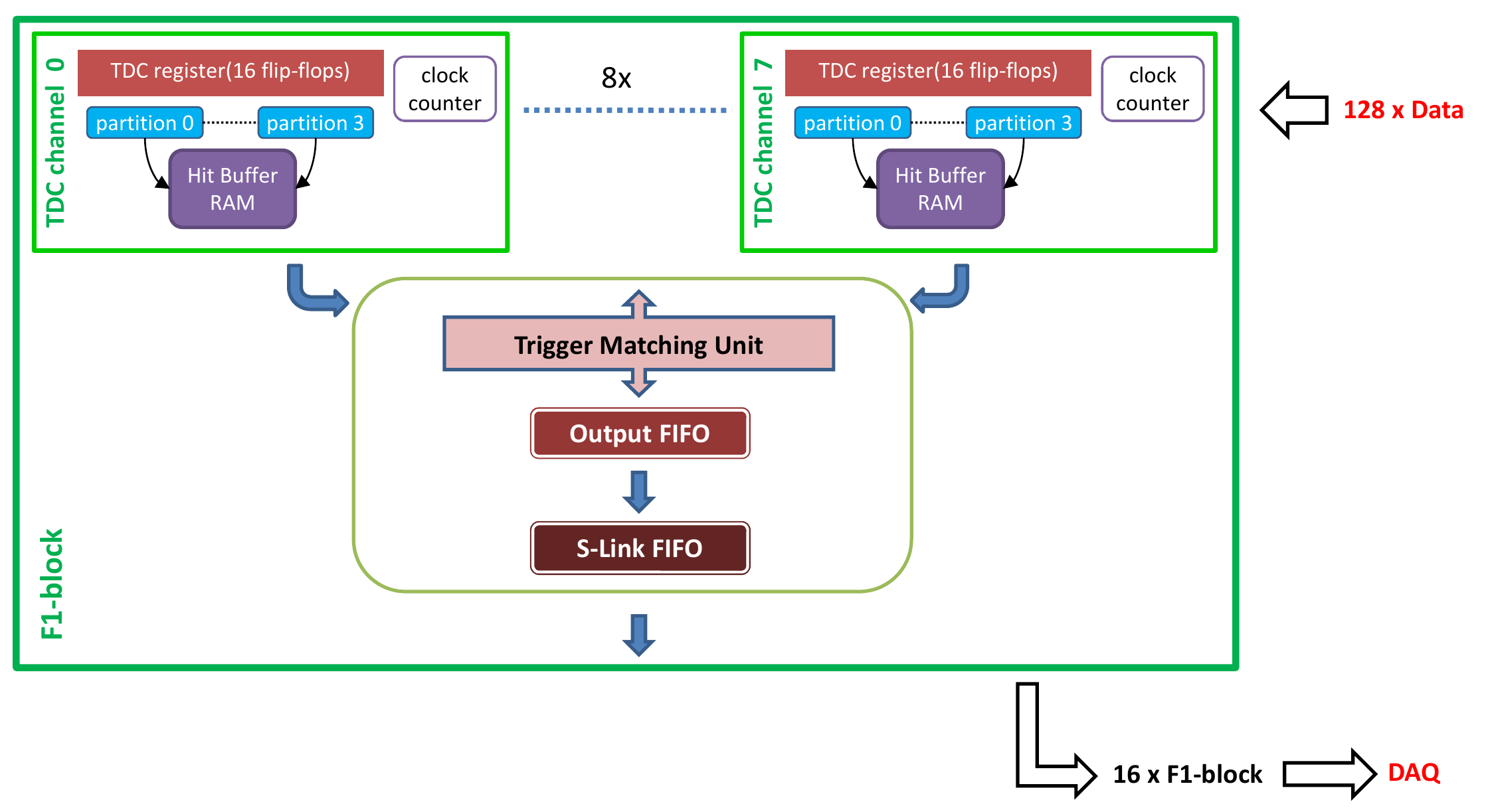}
	\caption{A F1-block combines 8 TDC channels into one common output FIFO.}
	\label{fig_f1block}
\end{figure}

The 128-channel TDC design is segmented into 16 identical blocks of 8 channels each, the so called 'F1-blocks' (Fig. \ref{fig_f1block}). This is done to ease the data collection process by using a two-step procedure. The timestamps of the detected hits are stored in a 1k deep hit buffer per channel. The timestamps of incoming triggers are buffered in a trigger FIFO, until they are processed by the trigger matching unit. This algorithm combines 8 channels at a time by selecting the hits from the respective hit buffers that fall into the trigger window and writing them to the output FIFO of the F1-block. Old hits beyond the trigger latency are deleted from the hit buffers. In a last step the data from all 16 F1-blocks are collected and sent to the DAQ using the S-Link interface. Thanks to the segmentation into F1-blocks, it was possible to use the same data output format as the existing hardware based on the F1 TDC chip \cite{fischer-f1,fischer}.

\subsection{FPGA Implementation}
To achieve good linearity in the digitization process, the TDC bin widths have to be as uniform as possible. The main contributions to the bin width variations are the clock phase error and the routing skew of the data signal to the different flip-flops. The first is very well controlled by the operation of clocking resources available in the FPGA. Two PLLs are used to generate 8 phase-shifted clocks that are distributed over the FPGA via global clock nets. For the 16-bin TDC design, each clock is inverted locally inside the slices to generate 8 additional clocks.
The routing skew is more difficult to control, because the FPGA implementation tools have no handle to influence the router to choose certain connections. Timing constraints that are available for traditional FPGA logic only ask for a maximum delay but not for a certain value. Hence, the placement of the TDC flip-flops was controlled by user-defined scripts in a way that the auto-router inevitably finds appropriate connections.

The design was floorplanned by defining area constraints for every F1-block and fixing their positions, to support the place and route process. The implementation was carried out separately for each F1-block and the results were saved as design partitions. These partitions were imported in the final implementation run where the remaining data merging and interface logic was added. The final design uses 43\% of the flip-flops and 27\% of the LUTs available in the Virtex-5 SX95T.

\section{Measurement Results}

To characterize the time-to-digital converter we developed a pattern generator using the GANDALF hardware with LVDS output cards. It generates test pulses for 128 channels with variable delay and repetition rate. A test setup with 2 pattern generators, 2 GANDALF TDC modules, a trigger control system (TCS) \cite{TCS} and a DAQ with S-Link readout was installed. For the measurements we used a clock frequency of 388.8 MHz, which results in a TDC bin width of 160 ps.

\begin{figure}[!t]
\centering
\includegraphics[width=0.7\textwidth]{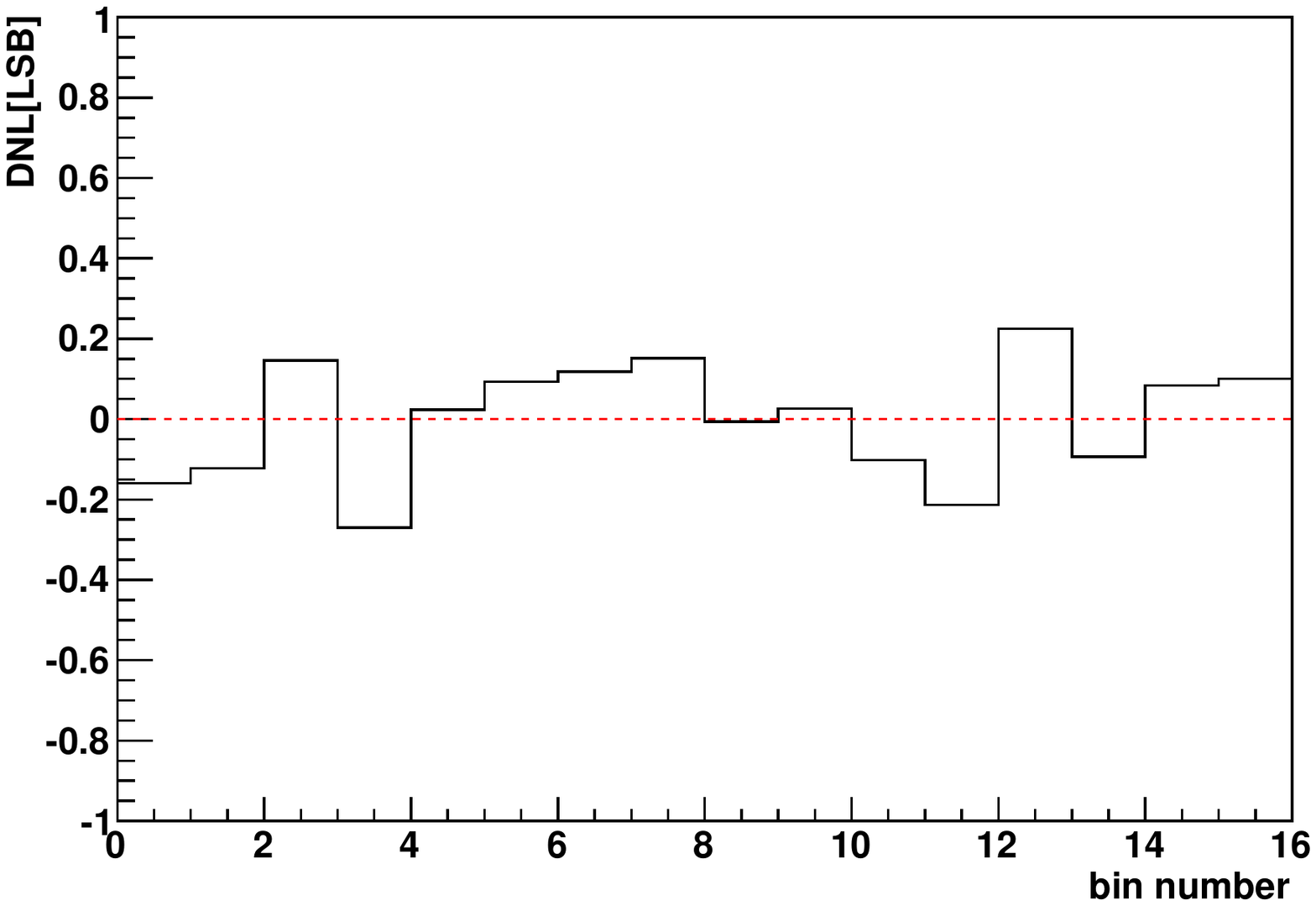}
\caption{Differential nonlinearity (DNL) of an exemplary TDC channel. The other channels show similar values.}
\label{fig_meas_dnl}
\end{figure}

\begin{figure}[!t]
\centering
\includegraphics[width=0.7\textwidth]{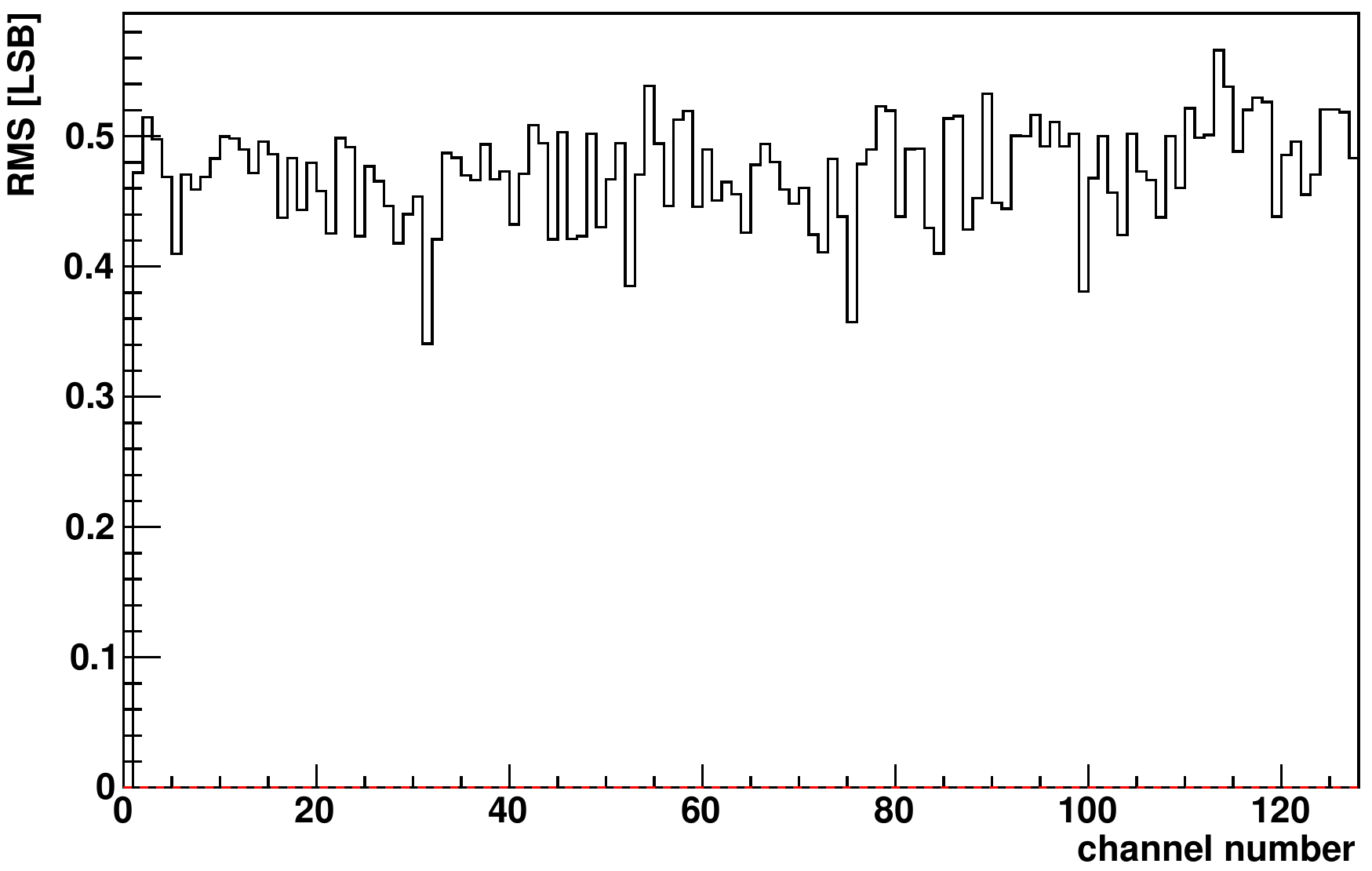}
\caption{RMS of the delay measurement vs. channel number. The RMS is below 0.56 LSB for all channels. For details see text.}
\label{fig_meas_rms}
\end{figure}

The differential nonlinearity (DNL) is a measure for the deviation of the TDC bin width from the nominal value. It was determined using code density tests with random pulses. Fig. \ref{fig_meas_dnl} exemplarily shows the result of the measurement for one channel.
The time resolution was determined by measuring the delay between two channels for a large number of events. Fig. \ref{fig_meas_rms} shows the RMS of the delay measurement for all channels. The RMS was divided by $\sqrt{2}$ to obtain the resolution for a single time measurement. This results in a TDC resolution better than $0.56 \cdot 160\:\mathrm{ps}\:\: / \sqrt{2} = 64\:\mathrm{ps}$.

\section{Conclusion and Outlook}
A 128-channel time-to-digital converter based on the shifted clock sampling method has successfully been implemented in a single Virtex-5 FPGA on the GANDALF module. 
The TDC base clock has a frequency of 388.8\:MHz and is divided into 16 TDC bins of 160\:ps each. The time resolution has been determined to 64\:ps.
With 43\% of the available flip-flops and 27\% of the available LUTs the device utilization of the current design is quite moderate, which allows for future extensions. At the moment work is ongoing to integrate 128 scaler channels into the same design for simultaneous rate measurements. Inter-board communication via the VXS interface is planned for fast trigger decisions.

\section*{Acknowledgements}
We gratefully acknowledge the discussions with our colleagues from the COMPASS collaboration and the support of the Freiburg workshops. 
The developments described in this report are supported by the German Bundesministerium f\"ur Bildung und Forschung and the European Community Research Infrastructure Integrating Activity under the FP7 Study of Strongly Interacting Matter (HadronPhysics2, Grant Agreement number 227431).





\bibliographystyle{elsarticle-num}
\bibliography{my-bib}







\end{document}